# *Proca* equation for laser pulses interaction with matter

**Janina Marciak- Kozłowska, Mirosław Kozłowski***
**Institute of Electron Technology, Al. Lotników 32/46, 02-668 Warsaw, Poland**

*Corresponding author

## Abstract

In this paper the interaction of ultrashort laser pulses with matter is investigated. The scattering and potential motion of heat carriers, as well as the external force are considered. It is shown that the heat transport is described by the *Proca* equation. For thermal Heisenberg type relation $V\tau \sim \hbar$, ($\tau$ is the relaxation time and $V$ is the potential) the solution of the Proca equation (PR) are the distortionless damped wave equation.

# 1 Introduction

Dynamical processes are commonly investigated using laser pump-probe experiments with a pump pulse exciting the system of interest and a second probe pulse tracking its temporal evolution. As the time resolution attainable in such experiments depends on the temporal definition of the laser pulse, pulse compression to the attosecond domain is a recent promising development.

After the standards of time and space were defined the laws of classical physics relating such parameters as distance, time, velocity, temperature are assumed to be independent of accuracy with which these parameters can be measured. It should be noted that this assumption does not enter explicitly into the formulation of classical physics. It implies that together with the assumption of the existence of an object and really independently of any measurements (in classical physics) it was tacitly assumed that *there was a possibility of an unlimited increase in the accuracy of measurements.* Bearing in mind the "atomicity" of time i.e. considering the smallest time period, the Planck time, the above statement is obviously not true. Attosecond laser pulses we are at the limit of laser time resolution.

With attosecond laser pulses belong to a new Nano-World where size becomes comparable to atomic dimensions, where transport phenomena follow different laws from that in the macro world. The first stage of miniaturization, from $10^{-3}$ m to $10^{-6}$ m is over and the new one, from $10^{-6}$ to $10^{-9}$ just beginning. The Nano-World is a quantum world with all the predicable and non-predicable (yet) features.

In this paper, we develop and solve the quantum relativistic heat transport equation for Nano-World transport phenomena where external forces exist. This is the generalisation of the results of paper [1] in which the quantum relativistic hyperbolic equation was proposed and solved. In this paper we describe the *Proca* thermal equation for an ultra–short laser pulse interaction with matter. A solution of the *Proca* equation is obtained for the *Cauchy* boundary conditions.

# 2 *Proca* thermal equation

Electromagnetic phenomena in vacuum are characterised by two three dimensional vector fields, the electric and magnetic fields *E(x,t)* and *B(x,t)* which are subject to Maxwell's equations and which can also be thought of as the classical limit of a quantum mechanical description in terms of photons. The photon mass is ordinarily assumed to be exactly zero in Maxwell's electromagnetic field theory, which is based on gauge invariance. Maxwell's equations must be replaced by the *Proca* equation [5, 6] for a photon with a rest mass $\neq 0$. The *Proca* equation reduces to (1), in free space, for a vector electromagnetic potential of $A_\mu$.

$$\left(\Box + \mu_\gamma^2\right) A_\mu = 0,$$
$$\Box = \frac{1}{c^2}\frac{\partial^2}{\partial t^2} - \frac{\partial^2}{\partial x^2}. \tag{1}$$

This is essentially the Klein – Gordon equation for massive photons. The parameter $\mu_\gamma$ can be interpreted as the photon rest mass $m_\gamma$, with

$$m_\gamma = \frac{\mu_\gamma \hbar}{c}. \tag{2}$$

It is quite interesting that the *Proca* type equation can be obtained for thermal phenomena. In the following starting with the hyperbolic heat diffusion equation the *Proca* equation for thermal processes will be developed and solved.

In paper [1] the relativistic hyperbolic transport equation was developed:

$$\frac{1}{v^2}\frac{\partial^2 T}{\partial t^2} + \frac{m_0 \gamma}{\hbar}\frac{\partial T}{\partial t} = \nabla^2 T. \tag{3}$$

In equation (3) $v$ is the velocity of heat waves, $m_0$ is the mass of heat carrier and $\gamma$ - the Lorentz factor, $\gamma = (1-\frac{v^2}{c^2})^{-1/2}$. As was shown in paper [1] the heat energy (*heaton temperature*) $T_h$ can be defined as follows:

$$T_h = m_0 \gamma v^2. \tag{4}$$

Considering that $v$, the thermal wave velocity equals [1]

$$v = \alpha c \tag{5}$$

where $\alpha$ is the coupling constant for the interactions which generate the *thermal wave*
($\alpha = 1/137$ and $\alpha = 0.15$ for electromagnetic and strong forces respectively) The *heaton temperature* is equal to

$$T_h = \frac{m_0 \alpha^2 c^2}{\sqrt{1-\alpha^2}}. \tag{6}$$

Based on equation (6) one concludes that the *heaton temperature* is a linear function of the mass $m_0$ of the heat carrier. It is interesting to observe that the proportionality of $T_h$ and the heat carrier mass $m_0$ was observed for the first time in ultrahigh energy heavy ion reactions measured at CERN [2]. In paper [2] it was shown that the temperature of pions, kaons and protons produced in Pb+Pb, S+S reactions are proportional to the mass of particles. Recently, at the Rutherford Appleton Laboratory (RAL), the VULCAN laser was used to produce the elementary particles: electrons and pions [3].

In the present paper the forced relativistic heat transport equation will be studied and solved. In paper [4] the damped thermal wave equation was developed:

$$\frac{1}{v^2}\frac{\partial^2 T}{\partial t^2} + \frac{m}{\hbar}\frac{\partial T}{\partial t} + \frac{2Vm}{\hbar^2}T - \nabla^2 T = 0. \tag{7}$$

The relativistic generalization of equation (7) is quite obvious:

$$\frac{1}{v^2}\frac{\partial^2 T}{\partial t^2} + \frac{m_0 \gamma}{\hbar}\frac{\partial T}{\partial t} + \frac{2Vm_0\gamma}{\hbar}T - \nabla^2 T = 0. \tag{8}$$

It is worthwhile noting that in order to obtain a non-relativistic equation we put $\gamma = 1$.

When an external force is present $F(x,t)$ the forced damped heat transport is obtained instead of equation (8) (in the one dimensional case):

$$\frac{1}{v^2}\frac{\partial^2 T}{\partial t^2} + \frac{m_0 \gamma}{\hbar}\frac{\partial T}{\partial t} + \frac{2Vm_0\gamma}{\hbar^2}T - \frac{\partial^2 T}{\partial x^2} = F(x,t). \tag{9}$$

The hyperbolic relativistic quantum heat transport equation, (10), describes the forced motion of heat carriers which undergo scattering ($\frac{m_0\gamma}{\hbar}\frac{\partial T}{\partial t}$ term) and are influenced by the potential term ($\frac{2Vm_0\gamma}{\hbar^2}T$).

Equation (10) can be written as

$$\left(\Box + \frac{2Vm_0\gamma}{\hbar^2}\right)T + \frac{m_0\gamma}{\hbar}\frac{\partial T}{\partial t} = F(x,t),$$

$$\Box = \frac{1}{v^2}\frac{\partial^2}{\partial t^2} - \frac{\partial^2}{\partial x^2}.$$

(10)

We seek the solution of equation (11) in the form

$$T(x,t) = e^{-t/2\tau}u(x,t)$$ (11)

where $\tau = \hbar/(mv^2)$ is the relaxation time. After substituting equation (12) in equation (11) we obtain a new equation

$$\left(\Box + q^2\right)u(x,t) = e^{\frac{t}{2\tau}}F(x,t)$$ (12)

and

$$q^2 = \frac{2Vm}{\hbar^2} - \left(\frac{mv}{2\hbar}\right)^2$$ (13)

$$m = m_0\gamma$$ (14)

In free space i.e. when F(x,t) → 0 equation (13) reduces to

$$\left(\Box + q^2\right)u(x,t) = 0$$ (15)

which is essentially the free *Proca* equation, compare equation (1).

The *Proca* equation describes the interaction of the laser pulse with the matter. As was shown in paper[ 1 ] the quantisation of the temperature field leads to the *heatons* – quanta of thermal energy with a mass $m_h = \hbar/{\tau v_h^2}$ [1], where $\tau$ is the relaxation time and $v_h$ is the finite velocity for heat propagation. For $v_h \to \infty$, i. e. for $c \to \infty$, $m_h \to 0$. It can be concluded that in non-relativistic approximation ($c$ = infinite) the *Proca* equation is the diffusion equation for mass less photons and heatons.

### 3.  Solution of the *Proca* thermal equation

For the initial Cauchy condition:

$$u(x,0) = f(x), \quad u_t(x,0) = g(x)$$ (16)

the solution of the *Proca* equation has the form (for q>0) [7]

$$u(x,t) = \frac{f(x-vt)+f(x+vt)}{2} \qquad (17)$$
$$+ \frac{1}{2v}\int_{x-vt}^{x+vt} g(\varsigma) J_0\left[q\sqrt{v^2t^2-(x-\varsigma)^2}\right] d\varsigma$$
$$- \frac{\sqrt{q}vt}{2}\int_{x-vt}^{x+vt} f(\varsigma) \frac{J_1\left[q\sqrt{v^2t^2-(x-\varsigma)^2}\right]}{\sqrt{v^2t^2-(x-\varsigma)^2}} d\varsigma$$
$$+ \frac{1}{2v}\int_0^t \int_{x-v(t-t')}^{x+v(t-t')} G(\varsigma,t') J_0\left[q\sqrt{v^2(t-t')^2-(x-\varsigma)^2}\right] dt'd\varsigma.$$

where $G = e^{t/2\tau} F(x,t)$.

When $q < 0$ solution of *Proc*a equation has the form:

$$u(x,t) = \frac{f(x-vt)+f(x+vt)}{2} \qquad (18)$$
$$+ \frac{1}{2v}\int_{x-vt}^{x+vt} g(\varsigma) I_0\left[-q\sqrt{v^2t^2-(x-\varsigma)^2}\right] d\varsigma$$
$$- \frac{v\sqrt{-q}t}{2}\int_{x-vt}^{x+vt} f(\varsigma) \frac{I_1\left[-q\sqrt{v^2t^2-(x-\varsigma)^2}\right]}{\sqrt{v^2t^2-(x-\varsigma)^2}} d\varsigma$$
$$+ \frac{1}{2v}\int_0^t \int_{x-v(t-t')}^{x+v(t-t')} G(\varsigma,t') I_0\left[-q\sqrt{v^2(t-t')^2-(x-\varsigma)^2}\right] d\varsigma dt'.$$

When $q = 0$ equation (14) is the forced thermal equation

$$\frac{1}{v^2}\frac{\partial^2 u}{\partial t^2} - \frac{\partial^2 u}{\partial x^2} = G(x,t). \qquad (19)$$

On the other hand one can say that equation (20) is the distortion less hyperbolic equation. The condition $q = 0$ can be rewritten as:

$$V\tau = \frac{\hbar}{8}. \qquad (20)$$

The equation (21) is the analogous to the Heisenberg uncertainty relations. Considering equation (4) equation (21) can be written as:

$$V = \frac{T_h}{8}, \quad V < T_h. \qquad (21)$$

It can be stated that distortion-less waves can be generated, only if $T_h > V$. For $T_h < V$, i.e. when the "Heisenberg rule" is broken, the shape of the thermal waves is changed.

## 3    Conclusions

In this paper we developed the relativistic thermal transport equation for an attosecond laser pulse interaction with matter. It is shown that the equation obtained is the *Proca* equation, well known in relativistic electrodynamics for massive photons. As the *heatons* are massive particles the analogy is well founded. Considering that for an attosecond laser pulse the damped term in Eq. (11) tends to 1, the transport phenomena are well described by the *Proca* equation.

# References

[1] J. Marciak-Kozłowska, M. Kozłowski,
*Lasers in Engineering*, **11**, (2001), p. 259.
[2] I. G. Bearden et al.,
*Phys. Rev. Lett.* **78**, (1997), p. 2080.
[3] K. W. D. Ledingham and P. A. Norreys,
*Contemporary Physics*, **40**, (1999), p. 367.
[4] M. Kozłowski, J. Marciak-Kozłowska,
*Lasers in Engineering,* **8**, (1998), p. 11.
[5] A.Proca *Compt. Rend.,* **190**, (1937), p. 1377.
[6] Liang – Cheng Tu et al.,
*Rep. Prog. Phys.* **68**, (2005), p. 77
[7] E. Zauderer,
*Partial Diferential Equation of Applied Mathematics,* Second Edition, Wiley 1989.